\documentclass[preprint,12pt]{elsarticle}

\usepackage{graphicx}

\usepackage{amssymb}

\journal{Nuclear Physics A}

\begin{document}

\begin{frontmatter}



\title{Nuclear medium effects on the $\bar K^*$ meson}


\author{Laura Tolos$^1$, Raquel Molina$^2$, E. Oset$^2$ and A. Ramos$^3$}

\address{$^1$Instituto de Ciencias del Espacio (IEEC/CSIC) Campus Universitat Aut\`onoma de Barcelona, Facultat de Ci\`encies, Torre C5, E-08193 Bellaterra (Barcelona),  Spain}

\address{$^2$Instituto de F{\'\i}sica Corpuscular (centro mixto CSIC-UV),
Institutos de Investigaci\'on de Paterna, Aptdo. 22085, 46071, Valencia, Spain}

\address{$^3$Departament d'Estructura i Constituents de la Mat\`eria,
Universitat de Barcelona,
Diagonal 647, 08028 Barcelona, Spain}

\begin{abstract}
The $\bar K^*$ meson in dense matter is analyzed by means of a unitary approach
in coupled channels based on the local hidden gauge formalism. The $\bar K^*$
self-energy and the corresponding $\bar K^*$ spectral function in the
nuclear medium are obtained. We observe that the $\bar K^*$ develops a width in
matter up to five times bigger than in free space. We also estimate the
transparency ratio of the $\gamma A \to K^+ K^{*-} A^\prime$ reaction. This
ratio is  an excellent tool to detect experimentally modifications of the $\bar
K^*$ meson in dense matter. 

\end{abstract}


\end{frontmatter}



\section{Introduction}

For years vector mesons in nuclear matter have been a matter of interest tied to
fundamental aspects of QCD \cite{rapp,hayano,mosel}. The $\rho$, $\omega$ and $\phi$ mesons in dense matter have been the focus of many theoretical and experimental analysis. The latest experiments conclude that there is no mass shift for the $\phi$ and $\rho$ in \cite{na60,wood}.
With regard to the $\omega$ meson, the debate started in  \cite{nucl-th/0610067}  concerning the background subtraction and was finally decided by explicitly determining the sources of background that there is no evidence of a mass shift \cite{nanova}.

Nevertheless, up to our knowledge, the properties of vector mesons with strangeness in matter have not been discussed in the literature. The fact that the $\bar K^*$ cannot be detected with dileptons might be a reason for the experimental neglect. From the theoretical point of view, recently the $\bar K^*N$ interaction in free space has been studied in  Ref.~\cite{GarciaRecio:2005hy} using SU(6) spin-flavour symmetry, and within the hidden local gauge formalism for the interaction of vector mesons with baryons of the octet \cite{Oset:2009vf} and the decuplet \cite{sarkar}.

In this paper we study the $\bar{K}^*$ self-energy in dense matter \cite{tolos10} following the work of Ref.~\cite{Oset:2009vf} and in a similar way as done in Ref.~\cite{Ramos:1999ku,angels,Tolos:2006ny}. We observe a spectacular enhancement of the $\bar{K}^*$ width in the medium, up to five times the free value of 50 MeV. In order to test this scenario experimentally, we also estimate the transparency ratio for $\bar{K}^*$ production in the $\gamma~A \to K^+ \bar{K}^{*-}A'$ reaction.

\section{The $\bar K^* N$ interaction}

The interaction of the  $\bar K^*$ mesons with nucleons is
obtained from $t$-channel vector-meson exchange mechanisms. The needed
interactions among vector mesons in these processes are derived 
within the hidden-gauge formalism \cite{hidden1,hidden2,hidden3,hidden4}, which
leads to the following three-vector
vertex Lagrangian
\begin{equation}
{\cal L}^{(3V)}_{III}=ig\langle (V^\mu\partial_\nu V_\mu -\partial_\nu V_\mu
V^\mu) V^\nu\rangle
\label{l3V}\ ,
\end{equation}
where $V_\mu$ is the SU(3)
matrix of the vectors of the octet of the $\rho$ plus the SU(3) singlet.
On the other hand, we also need the Lagrangian for the coupling of vector
mesons to
the baryons
\cite{Klingl:1997kf,Palomar:2002hk}:
\begin{equation}
{\cal L}_{BBV} = g\left( \langle \bar{B}\gamma_{\mu}[V^{\mu},B]\rangle +
\langle \bar{B}\gamma_{\mu}B \rangle \langle V^{\mu}\rangle \right) ,
\label{lagr82}
\end{equation}
where $B$ is the SU(3) matrix of the baryon octet.
Then, we can construct the Feynman diagrams that lead to  the vector-baryon ($VB$) transitions, $VB
\to V^\prime B^\prime$.

We proceed as in  Ref.~\cite{Oset:2009vf}
by neglecting the three momentum of the external vectors versus the vector
mass, as similarly done  for chiral Lagrangians in the low energy
approximation. Thus, one obtains the transition potential: 
\begin{equation}
V_{i j}= - C_{i j} \, \frac{1}{4 f^2} \, \left( k^0 + k^\prime{}^0\right)
~\vec{\epsilon}\,\vec{\epsilon }\,^\prime , \label{kernel} 
\end{equation} 
where $f$ is the pion decay constant;
$k^0, k^\prime{}^0$ are the energies of the incoming and outgoing vector
mesons, respectively; $\vec{\epsilon}\,\vec{\epsilon }\,^\prime$ is the product of their
polarization vectors; and $C_{ij}$ are the channel coupling  coefficients
\cite{Oset:2009vf}.

In order to obtain the meson-baryon scattering amplitude, we solve 
the on-shell Bethe-Salpeter equation in coupled channels \cite{angels,ollerulf}
\begin{equation}
 T = [1 - V \, G]^{-1}\, V \ ,
\end{equation}
with $G$ being the vector-baryon loop
function. The loop is regularized
 taking a natural value of $-2$ for the subtraction constants
at a regularization scale $\mu=630$ MeV \cite{Oset:2009vf,ollerulf}. We also incorporate  the relatively large decay width of the $\rho$ and  $\bar K^*$ vector mesons (into $\pi\pi$ or $\bar K\pi$ pairs, respectively)
in the loop functions via the 
convolution of the $G$ function \cite{nagahiro}.

\begin{figure}[t]
\begin{center}
\includegraphics[width=0.7\textwidth,]{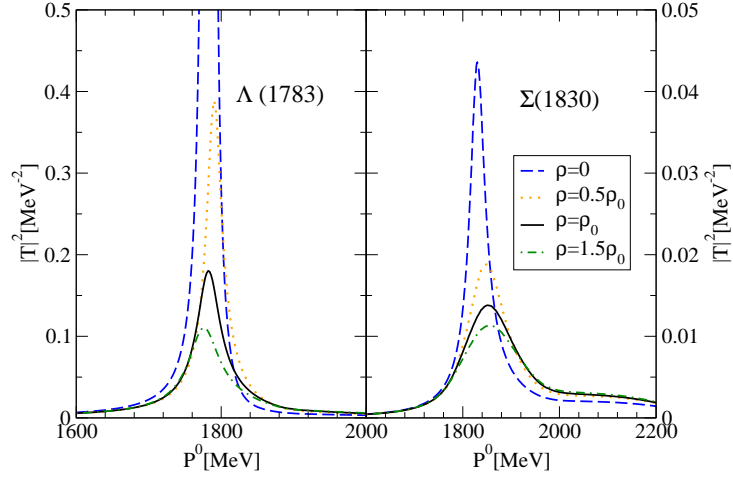}
\caption{($I=0$) $\Lambda(1783)$ and ($I=1$) $\Sigma(1830)$ states for various
densities, where $\rho_0=0.17$~fm$^{-3}$ is the nuclear
matter saturation density.}
\label{fig:reso}
\end{center}
\end{figure}

We analyze the vector meson-baryon $S=-1$ 
sector with isospin $I=0$ ($\bar K^*N$, $\omega \Lambda$, $\rho \Sigma$,
$\phi \Lambda$ and $K^* \Xi$) and $I=1$
($\bar K^*N$, $\rho \Lambda$, $\rho \Sigma$, $\omega \Sigma$, $K^* \Xi$ and
$\phi \Sigma$).  Two resonances in the $I=0$ and $I=1$ sectors are generated, 
 $\Lambda(1783)$ and  $\Sigma(1830)$ \cite{Oset:2009vf}, as displayed by the
dashed lines in Fig.~\ref{fig:reso}. These can be identified with the
experimentally observed states $J^P=1/2^-$ $\Lambda(1800)$ and $\Sigma(1750)$,
respectively, although the calculated  widths are
smaller than the ones measured experimentally due to the absence, in the present
approach, of the pseudoscalar-baryon decays in free space. Note that the
factorization of  the factor
$\vec{\epsilon}\,\vec{\epsilon }\,^\prime$ for the
external vector mesons appearing in the potential and also in the $T$ matrix \cite{Oset:2009vf} provides degenerate pairs of dynamically generated
resonances which have  $J^P=1/2^-,3/2^-$.

 Medium modifications on the $\bar K^* N$ scattering amplitude, which are
incorporated in the $\bar K^*N$ loop function, come from the Pauli blocking
effects acting on the nucleons and from the change in the properties of mesons
and baryons in nuclear matter. Specifically, we consider the self-consistent
$\bar K^*$ self-energy in the $\bar K^* N$ intermediate states, as explained
below. 
The modulus squared of the medium-modified $\bar K^* N$ amplitudes in
the isospin basis, which are obtained from the solution of the
on-shell Bethe-Salpeter coupled-channels equations in nuclear matter,
$T^{\rho,I} = [1 - V^I \, G^{\rho}]^{-1}\, V^I$, are displayed
in Fig.~\ref{fig:reso}, for $I=0$ and $I=1$, as functions of the
center-of-mass energy $P^0$ for a total momentum $\vec{P}=0$ and various
densities. Nuclear matter
effects result in a substantial increase of  their free widths due to the
opening of new decay channels in matter, such as $\bar K^* N \rightarrow \pi
\bar K N$, $\bar K^* N \rightarrow \bar
\pi \pi Y$, or $\bar K^* N N \rightarrow \bar K N N$.

\section{Contributions to the ${\boldmath \bar{K}^*}$ self-energy}


The $K^{*-}$ width in vacuum is determined by the imaginary part of the free $\bar K^*$ self-energy at rest,  ${\rm Im} \Pi^0_{\bar K^*}$, due to the decay of the $\bar{K}^*$ meson into $\bar{K}\pi$ pairs:  $\Gamma_{K^{*-}}=-\mathrm{Im}\Pi_{\bar{K}^*}^{0}/m_{\bar K^*}=42$ MeV \cite{tolos10}. Note that this value is quite close to the experimental value $\Gamma^{\rm exp}_{K^{*-}}=50.8\pm 0.9$ MeV.

\begin{figure}[t]
\begin{center}
\includegraphics[width=0.45\textwidth,height=5cm]{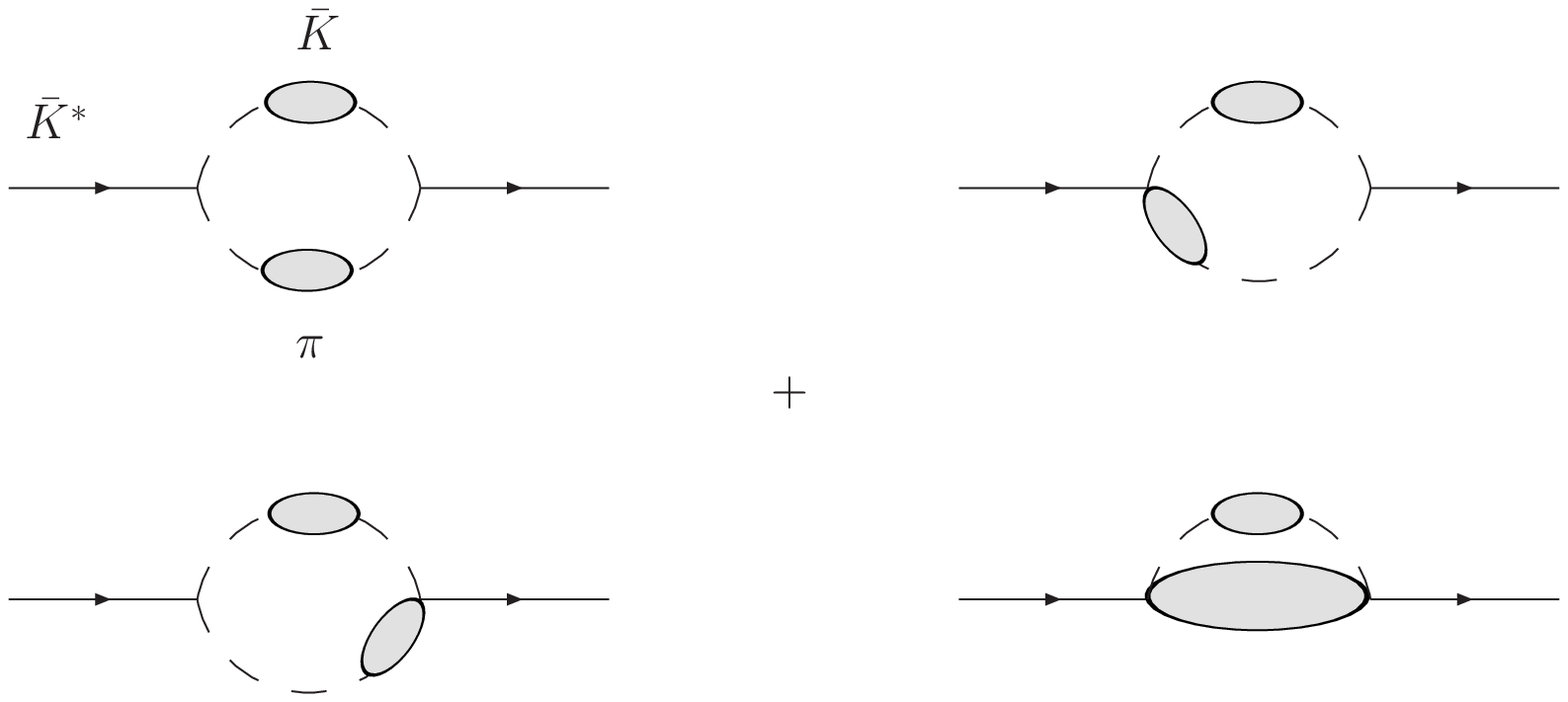}
\hfill
\includegraphics[width=0.5\textwidth,height=7cm]{Fig3_colouronline.eps}
\caption{Left figure: Self-energy diagrams from the decay of the $\bar{K}^*$ meson in the medium. Right figure: Imaginary part of the $\bar K^*$ self-energy at $\vec{q}=0 \, {\rm MeV/c}$, resulting from the ${\bar K}\pi$ decay in dense matter at  $\rho_0$.}
\label{fig:1}
\end{center}
\end{figure}

\begin{figure}[ht]
\begin{center}
\includegraphics[width=0.8\textwidth]{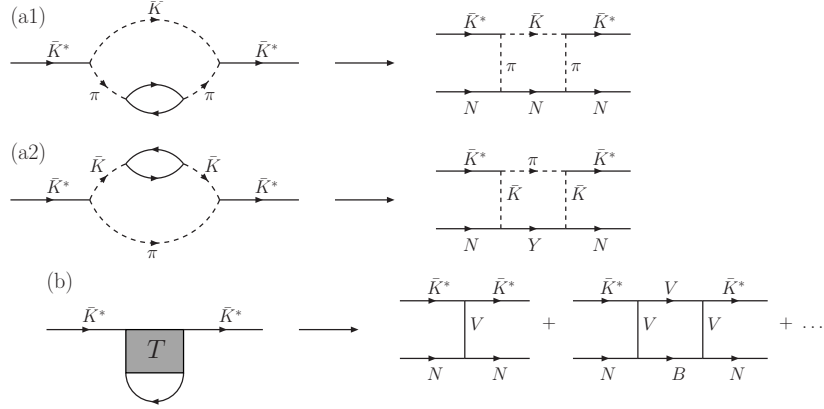}
\caption{Contributions to the $\bar K^*$ self-energy, depicting their different
inelastic sources.}
\label{fig:3}
\end{center}
\end{figure}

The $\bar{K}^*$ self-energy  in matter, on one hand, results from its decay into ${\bar K}\pi$, $\Pi_{\bar{K}^*}^{\rho,{\rm (a)}}$, including both the self-energy of the antikaon \cite{Tolos:2006ny} and the pion \cite{Oset:1989ey,Ramos:1994xy} (see first diagram on the left hand side of Fig.~\ref{fig:1} and some specific contributions in diagrams $(a1)$ and $(a2)$ of Fig.~\ref{fig:3}). Moreover, vertex corrections required by gauge invariance are also incorporated, which are associated to the last three diagrams in the l.h.s. of Fig. \ref{fig:1}.

We show the imaginary part of the $\bar{K}^*$ self-energy for $\vec{q}=0$ coming from $\bar{K}\pi$ decay in the right panel of Fig.~\ref{fig:1}:  free space calculation (dotted line),  adding the $\pi$ self-energy (dot-dashed line)
and including both $\pi$ and
$\bar{K}$  self-energy contributions (dashed line)  at
normal nuclear matter saturation density ($\rho_0=0.17$ \ fm$^{-3}$). We observe
that the main contribution to the $\bar K^*$ self-energy comes from the pion self-energy in dense matter. The reason is that the 
$\bar K^*  \to \bar K \pi$ decay process leaves the pion with energy right in
the region of $\Delta N^{-1}$ excitations, where there is considerable pionic
strength. The inclusion of vertex corrections (solid line) reduces the effect of the  pion dressing on the $\bar K^*$ self-energy, giving a $\bar K^*$ width of
$\Gamma_{\bar{K}^*}(\rho=\rho_0)=105$ MeV at the $\bar K^*$ mass. This turns out to be about twice the value of the width in vacuum.

The second contribution to the $\bar K^*$ self-energy comes from its interaction
with the nucleons in the Fermi sea, as displayed in diagram (b) of 
Fig.~\ref{fig:3}. This accounts for the direct quasi-elastic process $\bar K^* N \to \bar K^* N$ as well as other absorption channels $\bar K^* N\to \rho Y, \omega Y, \phi Y, \dots$ with $Y=\Lambda,\Sigma$. 
This contribution is determined by
integrating the medium-modified $\bar K^* N$ amplitudes, $T^{\rho,I}_{\bar
K^*N}$, over the  Fermi sea of nucleons,
\begin{eqnarray}
\Pi_{\bar{K}^*}^{\rho,{\rm (b)}}(q^0,\vec{q}\,)&=&\int \frac{d^3p}{(2\pi)^3} \, n(\vec{p}\,)\,
\left [~{T^\rho}^{(I=0)}_{\bar K^*N}(P^0,\vec{P})+3 {T^\rho}^{(I=1)}_{\bar K^*N}(P^0,\vec{P})\right ] \ ,
 \label{eq:pid}
\end{eqnarray}
where $P^0=q^0+E_N(\vec{p}\,)$ and $\vec{P}=\vec{q}+\vec{p}$ are the
total energy and momentum of the $\bar K^*N$ pair in the nuclear
matter rest frame, and the values $(q^0,\vec{q}\,)$ stand for the
energy and momentum of the $\bar K^*$ meson also in this frame. The
self-energy $\Pi_{\bar{K}^*}^{\rho,{\rm (b)}}$ has to be determined self-consistently
since it is obtained from the in-medium amplitude
${T}^\rho_{\bar K^*N}$ which contains the $\bar K^*N$ loop function
${G}^\rho_{\bar K^*N}$, and this last quantity itself is a function of the complete self-energy
$\Pi_{\bar K^*}^{\rho}=\Pi_{\bar{K}^*}^{\rho,{\rm (a)}}
+\Pi_{\bar{K}^*}^{\rho,{\rm (b)}}$.


We note that the two contributions to the $\bar K^*$ self-energy, coming from
the decay of
$\bar K \pi$ pairs in the medium [Figs.~\ref{fig:3}(a1) and \ref{fig:3}(a2)] or
from the  $\bar K^* N$ interaction [Fig.~\ref{fig:3}(b)] provide different
sources
of inelastic $\bar K^* N$ scattering, which add incoherently in the $\bar K^*$
width.  
As seen in the upper two rows of Fig.~\ref{fig:3}, the  $\bar K^* N$
amplitudes mediated by intermediate $\bar K N$ or $\pi
Y$ states are not unitarized. Ideally, one would like to treat the vector
meson-baryon ($VB$) and pseudoscalar meson-baryon ($PB$) states on the same
footing. However, at the energies of interest, transitions of the type $\bar K^*
N \to \bar K N$ mediated by pion exchange may place this pion on its mass shell,
forcing one to keep track of the proper analytical cuts contributing to the
imaginary part of the amplitude and making the iterative process more
complicated. A technical solution can be found by calculating the box diagrams
of Figs.~\ref{fig:3}(a1) and \ref{fig:3}(a2), taking all the cuts into account
properly, and adding the resulting $\bar K^* N \to \bar K^* N$ terms to the $VB
\to V^\prime B^\prime$ potential coming from vector-meson exchange, in a
similar way as done for the study of the vector-vector interaction in
Refs.~\cite{molinavec,gengvec}. Preliminar results of this procedure in free
space indicate that the generated resonances barely change their position for spin 3/2 and only by a moderate amount in some cases for spin 1/2. Their widths are somewhat enhanced due to the opening of the newly
allowed $PB$ decay channels \cite{garzonvec}.

\section{The $\bar K^*$ meson properties in dense matter}

\begin{figure}[ht]
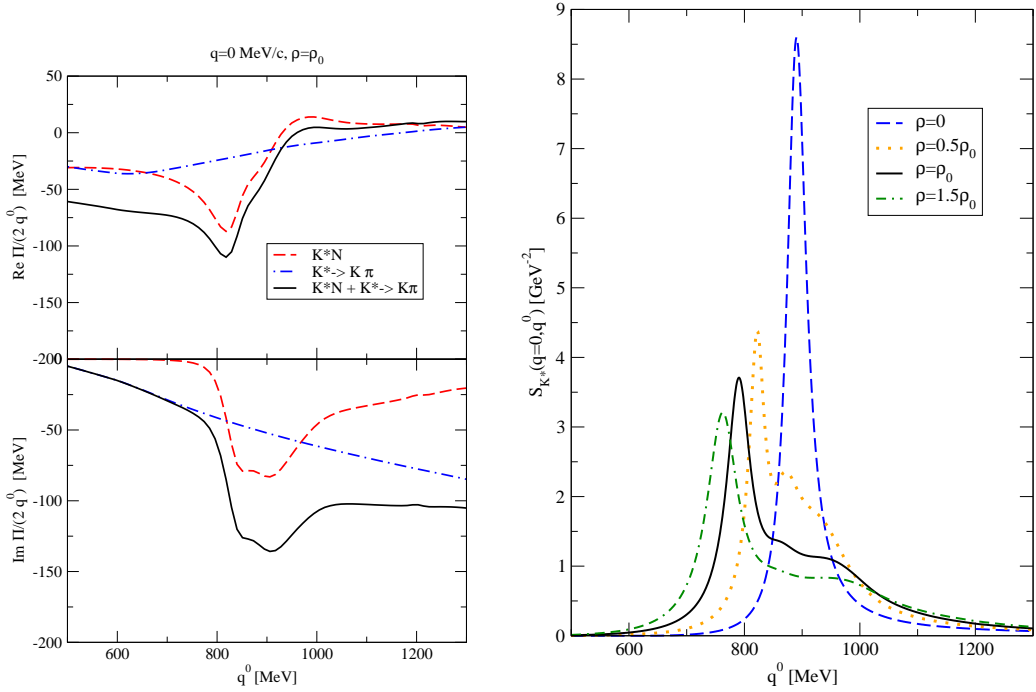

\begin{center}
\includegraphics[width=0.45\textwidth]{Fig5_colouronline.eps}
\hfill
\includegraphics[width=0.5\textwidth]{Fig7_colouronline.eps}
\caption{ Left figure: $\bar K^*$ self-energy for
 $\vec{q}=0 \, {\rm MeV/c}$ and $\rho_0$. Right figure: $\bar K^*$ spectral function for $\vec{q}=0 \, {\rm MeV/c}$  and different densities.}
\label{fig:auto-spec}
\end{center}
\end{figure}

In this section we show results for the $\bar K^*$ self-energy and the
corresponding
$\bar K^*$ spectral density in dense matter, which is obtained
from the imaginary part of
the in-medium $\bar K^*$ propagator, and is given by
\begin{equation}
S_{\bar K^*}(q^0,\vec{q}\,)=-\frac{1}{\pi} \, {\rm Im} \, \left[
\frac{1}{(q^0)^2-(\vec{q}\,)^2-m_{\bar K^*}^2-\Pi_{\bar
K^*}^{\rho}(q^0,\vec{q}\,)} \right] \ .
\end{equation}

The full $\bar K^*$ self-energy as a function of the $\bar K^*$ energy for zero
momentum at normal nuclear matter density is shown  in the left hand side of
Fig.~\ref{fig:auto-spec}. We explicitly indicate the contribution to the
self-energy coming from the self-consistent calculation of the $\bar K^* N$
effective interaction (dashed lines) and the self-energy from the $\bar K^*
\rightarrow \bar K \pi$ decay mechanism (dot-dashed lines), as well as the
combined result from both sources (solid lines).

Around $q^0= 800-900$ MeV we observe an enhancement of the width as well as some structures in the real part of the $\bar K^*$ self-energy. The origin of these structures can be traced back to  the coupling of the $\bar K^*$ to the in-medium $\Lambda(1783) N^{-1}$ and  $\Sigma(1830) N^{-1}$ excitations, which dominate the $\bar K^*$ self-energy in this energy region. However, at lower energies where the $\bar K^* N\to V B$ channels 
are closed, or at large energies beyond the resonance-hole excitations,
the width of the $\bar K^*$ is governed by the $\bar K \pi$ decay mechanism in dense matter.

The $\bar K^*$ meson spectral function is displayed in the right panel of Fig.~\ref{fig:auto-spec} as a function of the $\bar K^*$ meson energy, for zero momentum and different densities up to 1.5 $\rho_0$. The calculation in free space, where only the $\bar K \pi$ decay channel contributes, is given by the dashed lines while the other three lines correspond to fully self-consistent calculations, which also incorporate the process $\bar K^* \rightarrow \bar K \pi$ in the medium. 

 The  $\Lambda(1783) N^{-1}$ and  $\Sigma(1830) N^{-1}$ excitations are present at the right-hand side of the quasiparticle peak, which is given by
 \begin{eqnarray}
 \omega_{qp}(\vec{q}=0)^2=m^2+{\rm Re} \Pi_{\bar K^*} (\omega_{qp}(\vec{q}=0,\vec{q}=0) \ . 
 \end{eqnarray}
Nuclear matter effects result in a dilution and merging of those resonant-hole states as well as a general broadening of the spectral function  due to the increase of collisional and absorption processes. Thus, the $\bar K^*$ feels a moderately attractive optical potential and acquires a width of $260$ MeV, which is about five times its width in vacuum.

\section{Transparency ratio for  $\gamma A \to K^+ K^{*-} A'$}

The width of the $\bar K^*$ meson in nuclear matter can be analyzed experimentally by means of the nuclear transparency ratio. The aim is to compare the cross sections of the photoproduction reaction $\gamma A \to K^+ K^{*-} A'$ in different nuclei, and tracing the differences to the in medium $K^{*-}$ width.

\begin{figure}[ht]
\begin{center}
\includegraphics[width=0.6\textwidth]{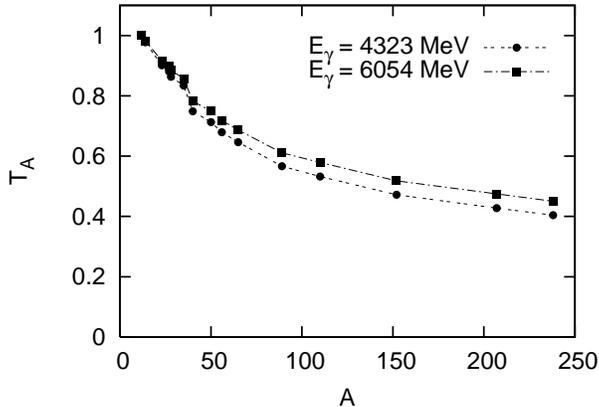}
\caption{Nuclear transparency ratio  for  $\gamma A \to K^+ K^{*-} A'$ for
different nuclei}
\label{fig:ratio}
\end{center}
\end{figure}

The normalized nuclear transparency ratio is defined as
\begin{equation}
T_{A} = \frac{\tilde{T}_{A}}{\tilde{T}_{^{12}C}} \hspace{1cm} ,{\rm with} \ \tilde{T}_{A} = \frac{\sigma_{\gamma A \to K^+ ~K^{*-}~ A'}}{A \,\sigma_{\gamma N \to K^+ ~K^{*-}~N}} \ .
\end{equation}
The quantity $\tilde{T}_A$ is the ratio of the nuclear $K^{*-}$-photoproduction cross section
divided by $A$ times the same quantity on a free nucleon. This describes the loss of flux of $K^{*-}$ mesons in the nucleus and is related to the absorptive part of the $K^{*-}$-nucleus optical potential and, therefore, to the $K^{*-}$ width in the nuclear medium.  In order to remove other nuclear effects not related to the absorption of the $K^{*-}$, we evaluate this ratio with respect to $^{12}$C, $T_A$.

In  Fig.~\ref{fig:ratio} we show the transparency ratio for different nuclei and for two energies in the center of mass reference system, $\sqrt{s}=3$ GeV and $3.5$ GeV, or, equivalently, two energies of the photon in the lab frame of $4.3$ GeV and $6$ GeV respectively. There is a very strong attenuation of the $\bar{K}^*$ survival probability coming from the decay or absorption channels $\bar{K}^*\to \bar{K}\pi$ and $\bar{K}^*N\to \bar K^* N, \rho Y, \omega Y, \phi Y, \dots$, with increasing nuclear-mass number $A$. The reason is the larger path that the $\bar{K}^*$ has to follow before it leaves the nucleus, having then more chances of decaying or being absorbed.

\section{Conclusions}

The properties of $\bar K^*$ mesons in symmetric
nuclear matter are studied within a self-consistent coupled-channel
unitary approach based on hidden-gauge local symmetry. We obtain the self-energy and, hence, the spectral
function of the $\bar K^*$ meson. The corresponding in-medium solution incorporates Pauli blocking
effects and the $\bar K^*$ meson self-energy in a
self-consistent manner, the latter one including the $\bar K^* \to \bar K \pi$ decay in dense matter.

We have observed an important change in
the $\bar K^*$ width in nuclear matter as compared to free space. At normal nuclear matter density the $\bar{K}^*$ width is found to be about $260$ MeV, five times larger than its free width. We have also made an estimate of the transparency ratio for different nuclei in the $\gamma A\to
K^+\bar{K}^* A'$ reaction. We have found a substantial reduction from unity of that
magnitude due to decay and absorptive new mechanisms in matter.  

\section*{Acknowledgments}
\vspace{0.5cm}

L.T. acknowledges support from the Ministerio de Ciencia y Tecnolog\'ia under  FPA2010-16963 contract and RyC2009 Programme,  from  FP7-PEOPLE-2011-CIG under PCIG09-GA-2011-291679 and the Helmholtz
 International Center for FAIR within the framework of the LOEWE
 program by the State of Hesse (Germany). 
 This work is partly supported by the EU contract No.
MRTN-CT-2006-035482 (FLAVIAnet), projects FIS2006-03438 and
FIS2008-01661 from the Ministerio de Ciencia e Innovaci\'on (Spain),
by the Generalitat Valenciana in the program Prometeo and  by the
Ge\-ne\-ra\-li\-tat de Catalunya contract 2009SGR-1289.  We acknowledge the support of the European Community-Research Infrastructure Integrating Activity ``Study of Strongly Interacting Matter'' (HadronPhysics2, Grant Agreement n. 227431) under the 7th Framework Programme of EU. 








\end{document}